\documentclass{Interspeech2024}
\usepackage{tipa}
\usepackage[symbol]{footmisc}

\interspeechcameraready

\title{Deep Speech Synthesis from Multimodal Articulatory Representations}

\name[affiliation={1}]{Peter}{Wu}
\name[affiliation={1}]{Bohan}{Yu}
\name[affiliation={2}]{Kevin}{Scheck}
\name[affiliation={3}]{Alan W}{Black}
\name[affiliation={1}]{Aditi S.}{Krishnapriyan}
\name[affiliation={1}]{Irene Y.}{Chen}
\name[affiliation={2}]{Tanja}{Schultz}
\name[affiliation={3}]{Shinji}{Watanabe}
\name[affiliation={1}]{Gopala K.}{Anumanchipalli}

\address{
  $^1$University of California, Berkeley\\
  $^2$University of Bremen\\
  $^3$Carnegie Mellon University}
\email{peterw1@berkeley.edu}


\begin{document}

\maketitle

\begin{abstract}
The amount of articulatory data available for training deep learning models is much less compared to acoustic speech data. In order to improve articulatory-to-acoustic synthesis performance in these low-resource settings, we propose a multimodal pre-training framework. On single-speaker speech synthesis tasks from real-time magnetic resonance imaging and surface electromyography inputs, the intelligibility of synthesized outputs improves noticeably. For example, compared to prior work, utilizing our proposed transfer learning methods improves the MRI-to-speech performance by 36\% word error rate. In addition to these intelligibility results, our multimodal pre-trained models consistently outperform unimodal baselines on three objective and subjective synthesis quality metrics.
\end{abstract}

\section{Introduction}

Articulatory synthesis incorporates information about the vocal tract into speech synthesizers to improve interpretability, generalizability, and efficiency \cite{browman1984articulatory, aryal2016data, wu2022emaspeech, wu2023mrispeech, krug23_interspeech}. Since these models are biologically grounded, they can be applied to decoding speech from biosignals for health technology applications \cite{csapo2017dnn, kimura2019sottovoce, gaddy2021improved, scheck2023emg_sil, metzger2023ecog, lian2023unconstrained, bocquelet2016real, luo2023brain}. These qualities make articulatory synthesis a promising direction for building speech synthesizers that support all types of speakers and perform well in low-resource settings.

However, there is much less articulatory data than other types of language data like acoustics and text \cite{richmond2011mngu0, Tiede2017HPRC, zen2019libritts}. Thus, current articulatory synthesizers output lower-fidelity speech than non-articulatory ones \cite{wu2022emaspeech, wu2023mrispeech, scheck2023emg_sil}. To improve the generalizability of articulatory synthesizers, prior work has studied appending acoustic- and articulatory-derived features to the input \cite{wu2022emaspeech}, modifying the model architecture \cite{gaddy2021improved, wu2022emaspeech}, fine-tuning from pre-trained weights \cite{wu2023mrispeech}, and training cross-speaker models \cite{Scheck24EMBC}.

Another way to improve articulatory synthesizers is multimodal learning, i.e., jointly training with multiple speech modalities \cite{liang2021multibench}. Each modality offers a different view of the same underlying speech representation space, and the unique pros and cons of each articulatory modality make them complementary. For example, there is more electromyography (EMG) data than electromagnetic articulography (EMA) data since EMG is easier to collect, but EMG is much noiser \cite{gaddy2021improved, richmond2011mngu0}. Real-time magnetic resonance imaging (MRI) scans of the vocal tract provide articulator position information at a much higher spatial density than EMA and EMG. However, audio recorded during MRI scans is much noiser than recordings with the EMA and EMG. We summarize several characteristics of these modalities in Table \ref{tab:data}.

We propose a multimodal pre-training approach for improving articulatory synthesis, validated through MRI- and EMG-to-speech tasks. With less than 10 minutes of single-speaker training data, our MRI-to-speech model achieves a test-set automatic speech recognition (ASR) word error rate (WER) of 33.4\%, compared to 69.5\% from the previous model \cite{wu2023mrispeech}. Our EMG-to-speech model similarly noticeably outperforms the baseline, and additional objective and human listening tests results match these trends.

\section{Related Work}

\subsection{Deep Articulatory Synthesis}
Deep articulatory synthesis involves synthesizing acoustics from articulatory features using deep learning \cite{aryal2016data, csapo2017dnn, kimura2019sottovoce, gaddy2021improved, wu2022emaspeech, wu2023mrispeech, scheck2023emg_sil, krug23_interspeech}. Current approaches can generally be described as either direct or involving an intermediate feature. Direct synthesis maps articulatory inputs to acoustics with a single end-to-end model \cite{wu2022emaspeech, wu2023mrispeech, Scheck23StreamETS}. On the other hand, synthesis with intermediates maps inputs to intermediate features, which are then mapped to acoustics \cite{gaddy2021improved, scheck2023emg_sil}. While prior work has proposed multiple models that can synthesize intelligible speech, these models focus on a single articulatory modality. In this paper, we extend the intermediate-based approach popular in EMG-to-speech synthesis \cite{gaddy2021improved, scheck2023emg_sil} to the multimodal setting.

\subsection{Multimodal Pre-Training}

Multimodal pre-training involves training a model with multiple modalities jointly, with the resulting model able to perform better in downstream tasks compared to models trained with fewer modalities \cite{liang2021multibench, wang2023large}. Multimodal speech processing has seen success in many tasks, from emotion recognition to text-to-speech, but utilizing multimodal learning to improve articulatory synthesis remains underexplored \cite{zhao2013real, tamura2015audio, yoon2020speech, zhang2021talnet, wang2023multimodal}. \cite{carlson2005data} identifies the usefulness of multimodal inputs for articulatory synthesis, and we explore this idea in the context of deep learning, which current state-of-the-art methods use. \cite{chen2021ema2s} describes a speech synthesizer with EMA and spectrogram inputs, providing a deep architecture for multimodal articulatory synthesis. Since they experiment with only one articulatory modality, EMA, we study whether jointly training with multiple articulatory modalities improves synthesis quality. We also introduce a simpler model that omits their deep feature training loss but still noticeably improves performance. The deep feature loss requires datapoints to have labels for multiple modalities. Since most articulatory datasets only label one articulatory modality \cite{richmond2011mngu0, Tiede2017HPRC, udupa2021estimating, gaddy2021improved, scheck2023emg_sil, lim2021multispeaker}, omitting this loss term enables us to do multimodal learning with these unimodal datasets.

\section{Methods}

We propose an encoder-decoder-based framework for multimodal articulatory synthesis. Like \cite{gaddy2021improved}, the encoder and decoder are trained separately, since the former requires articulatory labels whereas the latter does not. Our encoder utilizes a multimodal fusion layer to jointly encode multiple modalities, and our decoder is an articulatory vocoder that maps encoder outputs to waveforms end-to-end, detailed below.

\subsection{Multimodal Encoder}
\label{sec:encoder}

Our encoder accepts multiple articulatory modalities as input and outputs a deep representation that is inputted to the decoder. Encoder inputs and outputs are sequences of vectors, and inputs all have the same sampling rate and are temporally aligned with outputs. Given modalities $\{1, 2, \dots, M\}$, for each modality $m \in [M]$, its datapoint $x_m \in \{x_{mi}\}_i$ is fed into unimodal encoder $e_m$. Then, $\{e_m(x_m)\}_m$ is fed into pooling layer $f$, followed by multimodal encoder $g$. Our encoder can thus be written as

\begin{equation}
e(\{x_m\}_m) = g(f(\{e_m(x_m)\}_m)).
\end{equation}

\begin{figure}[t]
\begin{center}
\includegraphics[width=60mm]{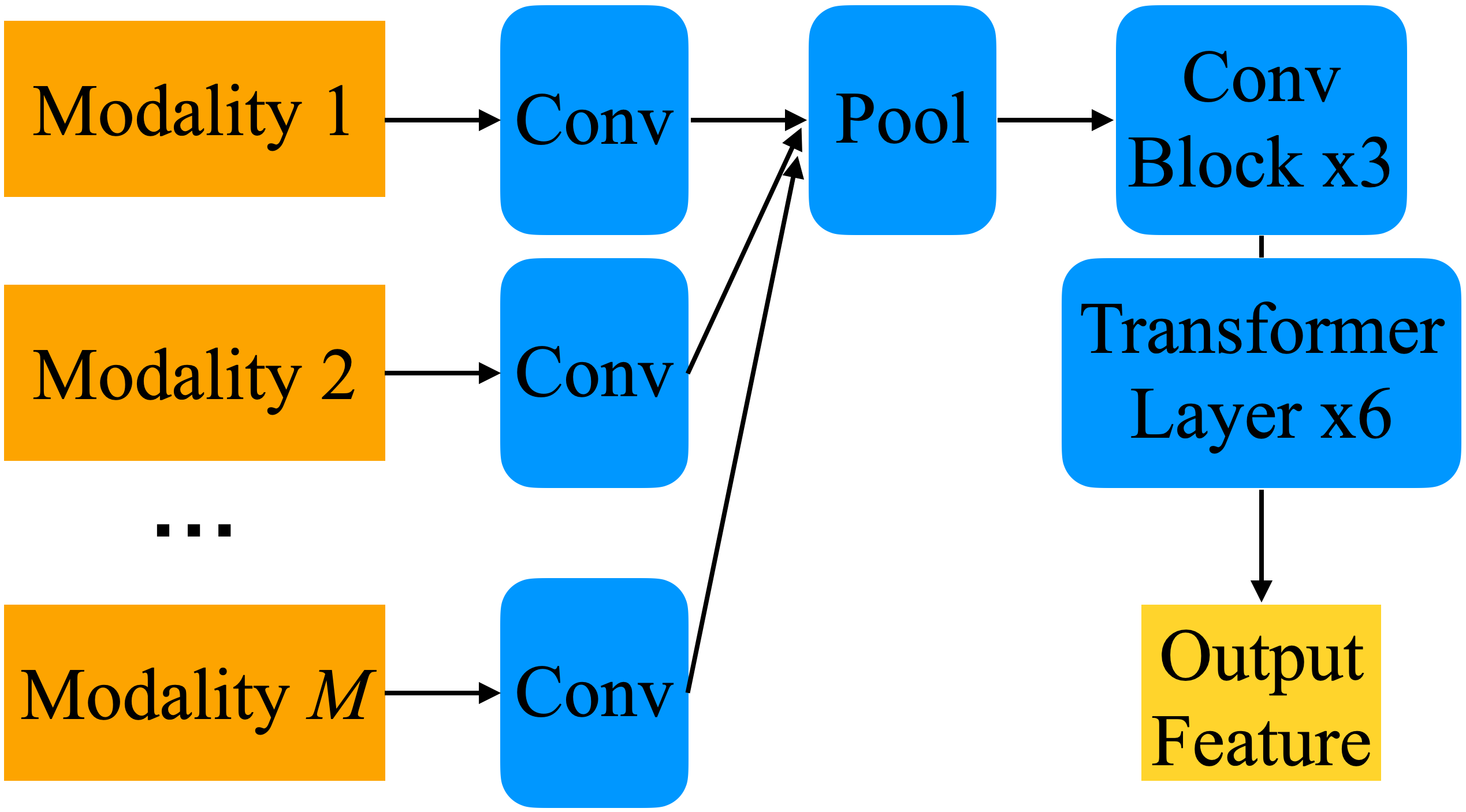}
\end{center}
\caption{Multimodal encoder architecture.}
\label{fig:encoder}
\vspace*{-3pt}
\end{figure}

Our encoder is visualized in Figure \ref{fig:encoder}. We build on the EMG-to-spectrum model in \cite{gaddy2021improved}, setting $g$ to be three residual convolution blocks \cite{he2016deep} followed by a six-layer Transformer \cite{vaswani2017attention}. Each residual block consists of three convolutions followed by a ReLU non-linearity \cite{fukushima1975cognitron}, as detailed in \cite{gaddy2021improved}. Our unimodal encoders are one-dimensional convolutions for simplicity, given the linear relationships between modalities in Section \ref{sec:modality_relationship}. We train the encoder using multimodal pre-training followed by fine-tuning with a single articulatory modality, detailed in Section \ref{sec:multimodal_pretrain}.

\subsection{Multimodal Fusion}
\label{sec:fusion}

We design our pooling layer $f$ such that our encoder output is invariant with respect to the choices of modalities inputted. For unimodal articulatory input $x_n$, we let $x_m = \vec{0}$ for all $m \neq n$. For conciseness, we write $e(\{x_m\}_m)$ in this case as $e(x_n)$. Defining

\begin{equation}
f(Z) = \sum_{z \in Z} z
\end{equation}

and $e_m(\vec{0}) = \vec{0}$, $\forall m \in [M]$, e.g., convolution with $\vec{0}$ bias, yields

\begin{equation}
e(x_n) = g(e_n(x_n)).
\end{equation}

This is desirable as unimodal encoders besides $e_n$ should not contribute to the encoder output when $n$ is the only input modality, motivating our architecture choice.

One optimality condition for our encoder is $e_m(x_m) = e_n(x_n)$ given $x_m$ and $x_n$ are representative of the same datapoint $x$, $\forall m,n \in [M]$, as reflected in the deep feature loss in \cite{chen2021ema2s}. By extension, modality invariance suggests that different views of the same datapoint should have the same encoding $e(\cdot)$. In other words, zeroing out different subsets of $\{x_m\}_m$ should yield the same $e(\{x_m\}_m)$. To satisfy this condition, we scale $f$ based on the number of inputted modalities, i.e.,

\begin{equation}
f(Z) = (\sum_{z \in Z} z)/N,
\end{equation}

where $N = |X_N|$ and $X_N = \{x_m|x_m\neq\vec{0}\}_m$. Our multimodal encoder and fusion methods are compatible with prior loss functions requiring more than one modality as input \cite{germain2018speech, chen2021ema2s}. Moreover, our approach supports settings where only one input modality is available per datapoint, as is often the case \cite{Tiede2017HPRC, lim2021multispeaker, gaddy2021improved}, elaborated in Section \ref{sec:multimodal_pretrain}.

\subsection{Decoder}
\label{sec:decoder}

Given the output of the encoder as input, our decoder outputs an acoustic waveform. Our decoder architecture is HiFi-CAR, an auto-regressive temporal convolutional network optimized with adversarial training \cite{wu2022emaspeech, su2020hifi, morrison2021chunked}. HiFi-CAR is composed of convolution blocks and an autoregression module. The convolution blocks alternate between transposed convolutions that upsample their inputs and residual blocks \cite{he2016deep} that can perform more complex transformations \cite{su2020hifi}. The autoregression module is a feedforward model that encodes the last chunk of generated audio in order to improve phase prediction \cite{morrison2021chunked}. We refer the reader to \cite{su2020hifi}, \cite{morrison2021chunked}, and \cite{wu2022emaspeech} for full details and code. Similarly to \cite{wu2023mrispeech}, we fine-tune the decoder from pre-trained spectrum-to-waveform vocoder weights to improve generalizability. Like other speech synthesis tasks where this model has been successfully applied \cite{wu2022emaspeech, metzger2023ecog}, our decoder inputs and outputs are temporally aligned. Like spectrum-to-waveform vocoders \cite{su2020hifi, morrison2021chunked}, our decoder inputs can be extracted from waveforms. Thus, our decoder training input-output pairs can be generated from any speech corpus. This avoids the articulatory data scarcity challenge, letting us only need multimodal pre-training for training the encoder.

\subsection{Multimodal Pre-Training and Unimodal Fine-Tuning}
\label{sec:multimodal_pretrain}

During multimodal pre-training, we train the encoder on a dataset with multiple articulatory modalities. For each input datapoint $x = \{x_m\}_m$, we set $x_m = \vec{0}$ for all modalities $m \in [M]$ that are not present in the ground truth. We optimize with the L1 loss function, given by $L_1(x, y) = |e(x) - y|$, where $e(x)$ and $y$ are the predicted and ground-truth outputs, respectively. Based on \cite{germain2018speech, chen2021ema2s}, we also experiment with adding a deep feature loss to $L_1$ to align unimodal encoder outputs:

\begin{equation}
\label{eqn:deep}
    L_2(x, y) = \left( \sum_{\substack{m, n \in X_N \\ m < n}} |e_m(x_m) - e_n(x_n)| \right) / \binom{N}{2},
\end{equation}
dividing by $\binom{N}{2}$ so that $L_2$ has the same scale for any $X_N$. After multimodal pre-training, we fine-tune the model on the low-resource articulatory modality of interest, e.g., MRI or EMG. Here, we also set the other inputted modalities to $\vec{0}$ and optimize with $L_1$.

\section{Results}

\subsection{Datasets}
\label{sec:data}

We study jointly training an articulatory synthesizer with electromagnetic articulography (EMA), real-time magnetic resonance imaging (MRI), and surface electromyography (EMG) datasets. We summarize characteristics of these three modalities in Table \ref{tab:data} and provide detailed descriptions of each dataset below.

\subsubsection{EMA Dataset}
\label{sec:ema_data}

EMA data is comprised of the midsagittal x-y coordinates of 6 articulatory positions: lower incisor, upper lip, lower lip, tongue tip, tongue body, tongue dorsum \cite{schonle1987ema, wu2022emaspeech}. We use the Haskins Production Rate Comparison database (HPRC), an 8-speaker dataset containing 7.9 hours of 44.1 kHz speech and 100 Hz EMA \cite{Tiede2017HPRC}. Waveforms are downsampled to 16 kHz to match the sampling rate of waveforms in other datasets used in this work. To maintain consistency with prior work \cite{wu2022emaspeech, wu2023speechema, siriwardena2023speechema}, we focus only on the midsagittal plane and discarded the provided mouth left and jaw left data in HPRC. We concatenate the 6 x-y coordinates to form a 12-dimensional vector at each time step. In addition to the 12-dimensional EMA data, like \cite{wu2022emaspeech}, we concatenate pitch to the input due to the lack of voicing information in EMA, forming a 13-dimensional vector input at each time step. We compute pitch using CREPE \cite{kim2018crepe}, extracting pitch at the EMA data sampling rate. Following \cite{scheck2023emg_sil}, we use HuBERT-Soft \cite{soft-vc-2022} as the decoder input feature. HuBERT-Soft is a 256-dimensional representation with a sampling rate of 50 Hz that we extract from 16 kHz waveforms using a pre-trained Transformer \cite{vaswani2017attention, soft-vc-2022}. We choose HuBERT-Soft due to prior results highlighting its suitability for speech synthesis over other deep representations \cite{soft-vc-2022}. The EMA data is randomly divided into an 85\%-5\%-10\% train-validation-set split, yielding 10,935, 643, and 1287 utterances, respectively.


\begin{table}[t]
\centering\caption{Summary of articulatory datasets in this work.\label{tab:data}}
\begin{tabular}{lccc}
\toprule
 & EMA & MRI & EMG \\
\midrule
Hours of Data        & $7.9$ & $0.2$ & $3.9$ \\
Noisy Audio          & False & True & False \\
Channels             & $12$ & $310$ & $8$ \\
Sampling Rate (Hz)   & $100$ & $83.\bar{3}$ & $1,000$ \\
Approx. Cost of Device (USD) & $1,000$ & $100,000$ & $1,000$ \\
\bottomrule
\end{tabular}
\end{table}

\subsubsection{MRI Dataset}
\label{sec:mri_data}

MRI provides a more comprehensive feature set of the human vocal tract than EMA \cite{lim2021multispeaker, wu2023mrispeech, otani23mri}. In addition to the six locations described by EMA, midsagittal MRI images contain locations of the hard palate, pharynx, epiglottis, velum, and larynx, all of which are useful for speech synthesis \cite{wu2023mrispeech}. We use the 11-minute, 236-utterance, single-speaker MRI dataset in \cite{wu2023mrispeech}. This dataset is comprised of $20$ kHz speech and $83.\bar{3}$ Hz midsagittal MRI data, with 170 x-y points annotated for each MRI frame. Following \cite{wu2023mrispeech}, we apply their deep speech enhancement technique to denoise target audio and discard annotated points on the back, reducing the number of points from 170 to 155. We use their 200-11-25 train-dev-test split, but omit single-word gibberish test set utterances unsuitable for the speech recognition metric in \cite{wu2023mrispeech}, yielding 12 test utterances. To match the sampling rates of other datasets, we upsample the MRI features to $100$ Hz and downsample the waveforms to $16$ kHz. Like our EMA dataset in Section \ref{sec:ema_data}, we extract 100 Hz pitch \cite{kim2018crepe} and 50 Hz HuBERT-Soft \cite{soft-vc-2022} features from waveforms. Prior work also shows that we can accurately estimate EMA from acoustics \cite{wu2023speechema, siriwardena2023speechema, udupa2021estimating, cho2023evidence}. Following \cite{cho2023evidence}, we train a linear regression model to map WavLM \cite{chen2022wavlm} features extracted from waveforms to the 12-dimensional EMA in Section \ref{sec:ema_data}. As in \cite{cho2023evidence}, we use the output of the 9th layer of WavLM, a 50 Hz 1024-dimensional feature that we linearly interpolate to 100 Hz to match the sampling rate of the EMA data. Since \cite{cho2023evidence} found that 20s of training data is sufficient, we also train on 2000 frames, randomly chosen from the HPRC train set in Section \ref{sec:ema_data} \cite{Tiede2017HPRC}. We use WavLM \cite{chen2022wavlm} and the trained regression model to extract EMA from waveforms in our MRI dataset, concatenating pitch to the regression output to form the 13-dimensional EMA feature in Section \ref{sec:ema_data}.

\subsubsection{EMG Dataset}
\label{sec:emg_data}

EMG measures electrical potentials caused by nearby muscle activity using electrodes placed on top
of the skin \cite{harris1962component}. When placed near articulators, EMG provides another low-dimensional manifold of articulatory movements \cite{harris1962component, denby2010emg, gaddy2021improved, scheck2023emg_sil}. We use the the 3.9-hour vocalized speech subset of the dataset in \cite{gaddy2021improved}, which consists of 8-dimensional EMG data and aligned speech, denoted ``Parallel Vocalized Speech" in \cite{gaddy2021improved}. \cite{gaddy2021improved} also has "silent" utterances which where the participant silently articulated the sentence without any vocalization. However, since our other articulatory modalities have aligned acoustics and silent EMG does not, we omit the silent subset in this work. Following \cite{scheck2023emg_sil}, Our train-dev-test data split contains 6,755, 199, and 98 utterances, respectively. Speech waveforms have a sampling rate of 16 kHz, and EMG 1000 Hz. Since raw EMG signals are noisy, these signals are commonly processed first before being used downstream \cite{gaddy2021improved}. \cite{gaddy2021improved} found processing EMG with three residual convolution blocks \cite{he2016deep} to be better than statistical methods. Thus, we also process EMG with their convolutional feature extractor, yielding 768-dimensional EMG features with a sampling rate of 100 Hz that are inputted to our encoder. To avoid spurious results due to overfitting, we make sure that our EMG train, dev, and test sets are subsets of the respective sets used to train the EMG feature extractor. We also extract 13-dimensional EMA features from the waveforms in this dataset using the WavLM-regression approach in Section \ref{sec:mri_data}.

\subsubsection{Decoder Dataset}
\label{sec:vctk}

We train our decoder in Section \ref{sec:decoder} with VCTK, which has 110 English speakers and a total of 44 hours of 44.1 kHz speech, randomly dividing speakers into an 85\%-5\%-10\% train-validation-test split \cite{veaux2017vctk}. We downsample audio to 16 kHz in order to match the sampling rate of other datasets in this work. Like the aforementioned datasets, we extract 50 Hz HuBERT-Soft \cite{soft-vc-2022} features from waveforms.

\subsection{Experimental Setup}
\label{sec:exp_setup}

For our encoder, described in Section \ref{sec:encoder}, Transformer layers have a hidden dimension of 768 and a dropout value of 0.2. The three residual blocks in the encoder have strides of 2, 1, and 1, in that order, in order to downsample the 100 Hz input to 50 Hz. Like \cite{gaddy2021improved}, convolutions in these blocks have a kernel size of 3. Our unimodal encoders have a kernel size of 5, stride 1, and padding 2 so that their input sequence lengths match the output lengths. When training the encoder, like \cite{wu2022emaspeech}, we randomly crop a 0.6 to 2.0 second window from each sample in the size-64 batch, with the window length fixed within the batch.

Our decoder, described in Section \ref{sec:decoder}, contains 4 upsampling blocks, each composed of a Leaky ReLU nonlinearity \cite{maas2013rectifier} followed by a transposed convolution like \cite{su2020hifi}. Transposed convolutions have strides of 8, 5, 4, and 2, in that order, in order to upsample 50 Hz inputs to 16000 Hz waveforms. Like \cite{su2020hifi}, each upsampling block is followed by 3 residual blocks with kernel sizes of 3, 7, and 11. Each residual block contains 3 dilated convolutions with dilations of 1, 3, and 5 \cite{su2020hifi}. Following \cite{wu2022emaspeech}, the autoregression module is a 5-layer feedforward network with Leaky ReLU nonlinearity \cite{maas2013rectifier} and input, hidden, and output dimensions of 512, 256, and 128, respectively. We also train the decoder with the random cropping approach used for the encoder, using a 0.16 to 1.0 second window and a batch size of 32. For pre-trained weights, we use the LibriTTS \cite{zen2019libritts} HiFi-GAN mel-spectrogram to speech vocoder weights in \cite{su2020hifi, wu2023mrispeech}. Since some decoder weights have different sizes than those of the pre-trained vocoder, we only load the weights with matching dimensions.

We apply our multimodal pre-training method to improving MRI-to-speech and EMG-to-speech models, omitting EMA-to-speech since existing models are already intelligible \cite{wu2022emaspeech}. For MRI-to-speech, we do multimodal pre-training with EMA and MRI, as well as with EMA, MRI, and EMG. Similarly, we pre-train with EMA + EMG and EMA + MRI + EMG for EMG-to-speech. For each modality subset experiment, we also study whether including the EMA features estimated from waveforms in Section \ref{sec:data} improves model performance. To compare with prior work, we use the HiFi-CAR (Section \ref{sec:decoder}) articulatory vocoder in \cite{wu2023mrispeech} as our MRI-to-speech baseline, which has upsample scales [8, 5, 3, 2] and directly maps $83.\bar{3}$ Hz MRI to $20$ kHz acoustics. Since our voiced EMG task does not have a baseline to our knowledge \cite{gaddy2021improved}, we also use HiFi-CAR, here with upsample scales [5, 4, 4, 2] to map the $100$ Hz EMG features in Section \ref{sec:emg_data} to $16$ kHz acoustics. All models are trained using the Adam optimizer \cite{kingma2015adam} with betas \{0.5, 0.9\} and a learning rate of $10^{-4}$.

\subsection{Metrics}

Following prior speech synthesis work, we evaluate models with both subjective and objective metrics \cite{hayashi2021espnet2, wu2022emaspeech}. For subjective evaluation, we use the popular 5-scale mean opinion score (MOS) test. Each model receives 100 unique ratings. Objective metrics include: (1) automatic speech recognition (ASR) performance for measuring intelligibility, (2) mel-cepstral distortion (MCD) \cite{fukada1992adaptive}, and (3) SpeechBERTScore \cite{saeki2024spbertscore}. ASR character error rate (CER) and word error rate (WER) is computed with the Whisper Large model \cite{radford2023robust}. MCD measures voice quality by comparing spectrums of generated speech to those of ground truth waveforms and is often used to measure speech synthesis performance \cite{wu2022emaspeech, saeki2024spbertscore}. SpeechBERTScore utilizes self-supervised representations to automatically estimate synthesis quality, correlating higher with human ratings compared to prior objective metrics \cite{saeki2024spbertscore}.

\begin{table}[b]
\centering
\caption{Correlation between linear regression and ground truth outputs for different speech features. Same and unavailable input-output pairs are omitted.\label{tab:regr}}
\begin{tabular}{ccccc}
\toprule
Input $\backslash$ Output & EMA & EMG & MRI & \cite{soft-vc-2022} \\
\midrule
EMA & - & 0.107 & 0.309 & 0.158 \\
EMG & 0.510 & - & - & 0.347 \\
MRI & 0.577 & - & - & 0.203 \\
\cite{soft-vc-2022} & 0.702 & 0.224 & 0.423 & - \\
\bottomrule
\end{tabular}
\end{table}

\begin{table*}[t]
\centering\caption{MRI-to-speech, training proposed model with tri-modal (EMA, MRI, EMG), bi-modal (EMA, MRI), and unimodal (MRI) data.\label{tab:asr_mri}}
\begin{tabular}{lccccc}
\toprule
Model & CER (\%) $\downarrow$ & WER (\%) $\downarrow$ & MCD$\downarrow$ & SpeechBERTScore$\uparrow$ & MOS$\uparrow$ \\
\midrule
Tri-modal Encoder-Decoder with Deep Feature Loss  & 26.427 & 41.326 & 8.3359 & 0.7305 & 3.30 $\pm$ 0.93 \\
Tri-modal Encoder-Decoder     & 21.361 & 39.005 & 8.3961 & 0.7257 & 3.11 $\pm$ 1.11 \\
Bi-modal Encoder-Decoder with Deep Feature Loss   & \textbf{20.098} & \textbf{33.434} & \textbf{8.1961} & \textbf{0.7390} & \textbf{3.48} $\mathbf{\pm}$ \textbf{0.86} \\
Bi-modal Encoder-Decoder           & 21.938 & 39.239 & 8.2466 & 0.7291 & 3.16 $\pm$ 1.08 \\
Unimodal Encoder-Decoder                & 30.988 & 50.865 & 8.4652 & 0.6956 & 3.06 $\pm$ 1.00 \\
Vocoder \cite{wu2023mrispeech} & 48.208 & 69.451 & - & - & - \\
\bottomrule
\end{tabular}
\end{table*}

\begin{table*}[t]
\centering
\caption{EMG-to-speech, training proposed model with tri-modal (EMA, MRI, EMG), bi-modal (EMA, EMG), and unimodal (EMG) data.\label{tab:asr_emg}}
\begin{tabular}{lcccccc}
\toprule
Model & CER (\%) $\downarrow$ & WER (\%) $\downarrow$ & MCD$\downarrow$ & SpeechBERTScore$\uparrow$ & MOS$\uparrow$ \\
\midrule
Tri-modal Encoder-Decoder with Deep Feature Loss  & 18.342 & \textbf{18.091} & \textbf{5.2336} & \textbf{0.8341} & \textbf{3.69} $\mathbf{\pm}$ \textbf{0.86} \\
Tri-modal Encoder-Decoder     & 17.240 & 18.866 & 5.3110 & 0.8323 & 3.63 $\pm$ 0.94 \\
Bi-modal Encoder-Decoder with Deep Feature Loss   & \textbf{16.537} & 19.616 & 5.3107 & 0.8339 & 3.67 $\pm$ 0.82 \\
Bi-modal Encoder-Decoder           & 16.542 & 21.021 & 5.3442 & 0.8284 & 3.46 $\pm$ 1.01 \\
Unimodal Encoder-Decoder                & 22.993 & 24.394 & 5.3800 & 0.8198 & 3.51 $\pm$ 0.96 \\
Vocoder \cite{wu2023mrispeech} & 40.445 & 43.561 & - & - & - \\
\bottomrule
\end{tabular}
\end{table*}

\subsection{Relationship Between Speech Representations}
\label{sec:modality_relationship}

We check the linear relationship between our different speech representations in order to gain intuition on the efficacy of training with them jointly. This is motivated by \cite{carlson2005data}, which found a high correlation between face- and tongue-based articulatory features and improved unit selection synthesis when utilizing both instead of one. Following the setup in \cite{cho2023evidence}, we train a linear regression model to map one speech feature to another and calculate the Pearson correlation between predicted and ground truth outputs on a held-out test set. We train on 2000 frames like \cite{cho2023evidence} and Section \ref{sec:ema_data}, and test on 200 frames randomly chosen from the remaining data. We train regression models for the following speech features: EMA, MRI, EMG, and HuBERT-Soft \cite{soft-vc-2022}. For experiments with EMG, we choose frames from the EMG dataset in Section \ref{sec:data}, and likewise for MRI. Since we lack a dataset with EMG-MRI pairs, we do not train a regression model mapping between these modalities. To map between HuBERT-Soft and EMA, we use the EMA dataset in Section \ref{sec:ema_data}.

Table \ref{tab:regr} summarizes our linear regression results. There is a high correlation between EMA estimated from HuBERT-Soft \cite{soft-vc-2022} and ground truth EMA, reflecting prior intelligible EMA-to-speech \cite{wu2022emaspeech} and SSL-EMA correlation \cite{cho2023evidence} results. The EMG-to-EMA and MRI-to-EMA correlation results are also reasonably high, suggesting that EMG and MRI share information with EMA and can benefit from joint training. Estimating HuBERT-Soft from these 3 articulatory modalities is difficult, reflecting the articulatory synthesis challenges faced in prior work \cite{wu2022emaspeech, wu2023mrispeech, gaddy2021improved, scheck2023emg_sil, zhang2021talnet, kimura2019sottovoce, csapo2017dnn}. Likewise, the difficulty of estimating EMG and MRI from HuBERT-Soft reflects prior articulatory inversion challenges \cite{sharma2021comparative, scheck2023ste}.

\subsection{Synthesis Quality}

Tables \ref{tab:asr_mri} and \ref{tab:asr_emg} summarize our MRI-to-speech and EMG-to-speech results, respectively. Multimodal pre-training results in much better ASR performance than the baseline for both MRI-to-speech and voiced EMG-to-speech. Our best MRI-to-speech WER, 33.4\%, is noticeably better than the 69.5\% WER from the previous model \cite{wu2023mrispeech}. Adding more modalities generally improves model performance across all of our metrics, with the largest improvement when going from unimodal to bi-modal training. Multimodal pre-training with estimated EMA features and our deep feature loss in Equation \ref{eqn:deep} also generally improves synthesis quality, suggesting that adding this loss term increases multimodal alignment.


\section{Conclusion}

We devise a multimodal pre-training framework for improving the performance of deep MRI- and EMG-to-speech models. Our MRI-to-speech synthesizer outperforms the test-set ASR WER of the previous model \cite{wu2023mrispeech} by 36\% WER, with our EMG-to-speech model similarly outperforming the baseline. On all of our objective and subjective synthesis quality metrics, our multimodal models outperform the unimodal ones, highlighting the effectiveness of multimodal pre-training. In the future, we are interested in extending these results to multi-speaker tasks and additional datasets.



\bibliographystyle{IEEEtran}
\bibliography{mybib}

\end{document}